\documentclass[doublecol]{epl2} 
\usepackage{amssymb}

\title{Packing of softly repulsive particles in a spherical box - a generalised Thomson problem}
\shorttitle{Title} 

\author{A. Mughal\inst{1}}
\shortauthor{F. Author \etal}

\institute{                    
  \inst{1} Institute of Mathematics and Physics, Aberystwyth University, Penglais, Aberystwyth, Ceredigion, Wales, SY23 3BZ, United Kingdom\\
}
\pacs{nn.mm.xx}{First pacs description}
\pacs{nn.mm.xx}{Second pacs description}
\pacs{nn.mm.xx}{Third pacs description}

\abstract{We study the (near or close to) ground state distribution of N softly repelling particles trapped in the interior of a spherical box. The charges mutually interact via an inverse power law potential of the form $1/r^\gamma$. We study three regimes in which the charges form an single spherical shell at the edge of the box ($\gamma=1$), a series of concentric shells of increasing density ($\gamma=2$) and  $\gamma=12$ for which the charges form shells with a more uniform charge distribution. We conduct numerical simulations for clusters containing up to 5000 charges and compare charge density across the system with continuum limit results. The agreement between numerical (discrete) results and the continuum limit is found to improve with increasing N.}

\begin{document}

\maketitle

\section{Introduction}

The {\it generalised} Thomson problem, as dubbed by Bowick et al. \cite{Bowick:2002}, is concerned with finding the minimal energy configuration of N  point particles, that are confined to the surface of a sphere, and repel each other via an inverse power law potential $\phi(\gamma)=1/r^{\gamma}$. For over a decade the study of such problems has yielded fundamental insights into crystallisation and order on curved surfaces. Applications include understanding virus morphology \cite{Zandi:2005}, self-assembly of colloids on emulsion droplets \cite{Bausch:2002} and multi-electron bubbles in superfluid helium \cite{Bowick:2006}.

However, the original Thomson problem has its origins in the ``plum'' pudding model of the atom. The model, proposed by J.J. Thomson in 1904, postulates that the atom consists of classical electrons embedded in a neutralising droplet of positively charged fluid \cite{Thomson:1904}. Although this pre-quantum era model of the atom is now obsolete, nevertheless, the Thompson problem and its variations continue to be of interest to modern science in areas such as packing problems \cite{Aste:2008}, for benchmarking various optimisation algorithms \cite{Cecka:2009} and as means of efficiently discretising space for lattice simulations \cite{huttig2008}.  

In this Letter we return to the original spirit of the Thomson problem and conduct numerical simulations to find (near) ground states for clusters of particles, in a spherical box of radius $R$, which interact through a potential $\phi(\gamma)$. Such inverse power law potentials have been widely studied (see  \cite{Bowick:2006}, \cite{Agrawal:1995}, \cite{Prestipino:2005}) and provide a continuous path between the hard sphere packing limit ($\gamma \rightarrow \infty$) and the soft one-component plasma ($\gamma = 1$). Such a path can be realised using dilute solutions of colloidal particles with polymer chains grafted onto their surface. The effective pairwise interaction between colloidal particles can be tuned by changing the thickness of the polymer layer  and modelled with an inverse-power-law potential of the type $\phi(\gamma)$ \cite{Prestipino:2005}, \cite{Likos:2001} .

We show that in the absence of a neutralising fluid, and in the limit of large N, the continuum limit ground state distribution of particles within the spherical box falls into three distinct regimes. (i) For $\gamma \leq 1$ the repulsive interaction is strong enough to drive all the particles to the surface of the sphere \cite{Levin:2003}, that is the problem of finding the ground state of N point-particles inside a sphere reduces to the problem studied by Bowick et al. (ii) While for $ 1 < \gamma < 3$ this is no longer the case and charges are found in the interior of the sphere with a non-uniform radial density. Finally, (iii) for $\gamma \geq 3$ the particles are distributed throughout the sphere with uniform density \cite{Hardin:2004}.

Although it is recognised that for $\gamma \geq 1$ the presence of charges in the interior of the spherical box becomes energetically favourable \cite{Levin:2003}, our simulations demonstrate that these internal charges also condense into spherical crystals (in which most charges have six nearest neighbours) giving rise to a series of concentric shells. Decreasing  the range of interaction (by increasing $\gamma$) drives particles into the interior and leads to a higher occupancy of the inner shells. This is in contrast to similar shell like structures observed in spherical dusty plasma crystals (so called Yukawa balls), where decreasing the range of interaction (achieved by decreasing the screening length) drives particles towards the exterior and leads to a higher occupancy of the outer shells \cite{baumgartner2009}. 

For $\gamma \geq 3$ the continuum limit charge density is  expected to be uniform. In finite sized clusters the charges are found to be self-organise into multi-shell arrangements which could in principle be compared with similar carbon-based structures such as nested Fullerines \cite{tomanek1993}. Although the precise morphology of these latter systems depends on quantum chemistry, nevertheless the approach described here may prove useful in distinguishing between simple features, which are largely geometric in nature, and more complex properties arising from chemistry. 

The ground state configuration of mutually repelling charges is of interest in the efficient discretisation of spaces as a framework for various numerical schemes. These include statistical sampling, finite element tessellations, quadrature, interpolation and as starting points for NewtonÕs method. In the case of a spherical geometry, in such problems the need can often arise for a greater degree of resolution either at the centre or towards the surface of the box. By fixing $\gamma$ to be in the range $ 1 < \gamma < 3$ the present study provides a simple method for generating a mesh with a varying radial point density.

The paper is organised as follows. We begin by detailing our numerical approach. We then elucidate the three regimes by considering three exemplary cases, these are: $\gamma=1,2$ and $12$. In each instance we first state or derive the expected continuum limit charge distribution and then compare this with a series of finite sized clusters. In addition we also provide a brief pictorial gallery that is representative of some of the low energy structures found by our numerical simulations.

\section{Numerical Approach}
The energy of a cluster of N particles, interacting via a potential $\phi(\gamma)$ and  confined to the interior of a sphere of radius R, by a hard wall potential is given by,
\begin{equation}
E
=
{\sum_i^N}V(r_i)
+
{\sum_{i<j}^N}\frac{1}{|{{\bf r}_i}-{{\bf r}_j}|^\gamma}, 
\label{eq:totalenergy}
\end{equation}
where 
\begin{eqnarray}
V(r_i)=\left\{ \begin{array}{ll}
0      & \mbox{for ${r_i}<R$} \\
\infty & \mbox{for ${r_i}\geq R$}
\end{array}
\right. 
\nonumber
\end{eqnarray}
and ${\bf r}_i=(r_i,\theta_i,\phi_i)$. Finding the global minimum for a function such as Eq. (\ref{eq:totalenergy}) is a difficult task. The number of metastable states proliferate exponentially	with N; consequentially the global minimum is obscured by a large number of local minima with energies close to that of the global minimum. There exist a number of heuristic methods for such problems. Although there is no guarantee of finding the global minimum, it is possible to find states close to it.

We found that the standard Metropolis simulated annealing algorithm to be more effective than a conjugate gradient algorithm. For a system with N particles the simulated annealing algorithm was run with typically $N \times (5 \times 10^6)$ Monte Carlo steps. The temperature of the simulation was decreased linearly. The average displacement of the charges at each temperature step was chosen by an automatic process to give an acceptance probability of $0.5 \pm 0.01$. Promising states were reheated and annealed repeatedly to iron out as many defects as possible. Finally the results were put through a conjugate gradient algorithm to remove any residual strains.

\section{ First case: $\gamma=1$}

For $\gamma \leq 1$ the inter particle repulsion is strong enough to drive all the charges to the edge of the spherical box. The Coulomb ($\gamma=1$) case is merely a reflection of the familiar result from electrostatics that, under static conditions, the charge density inside a conductor is always zero. Thus we expect the charges to be located on the surface of the sphere and for the charge density, in the continuum limit, to be described by a delta function of the form,
\begin{equation}
\rho(r)
=
\frac{N}{4\pi R^2}
\delta(r-R).
\label{eq:delta}
\end{equation}

\begin{figure*}
\begin{center}
\includegraphics[width=1.8\columnwidth ]{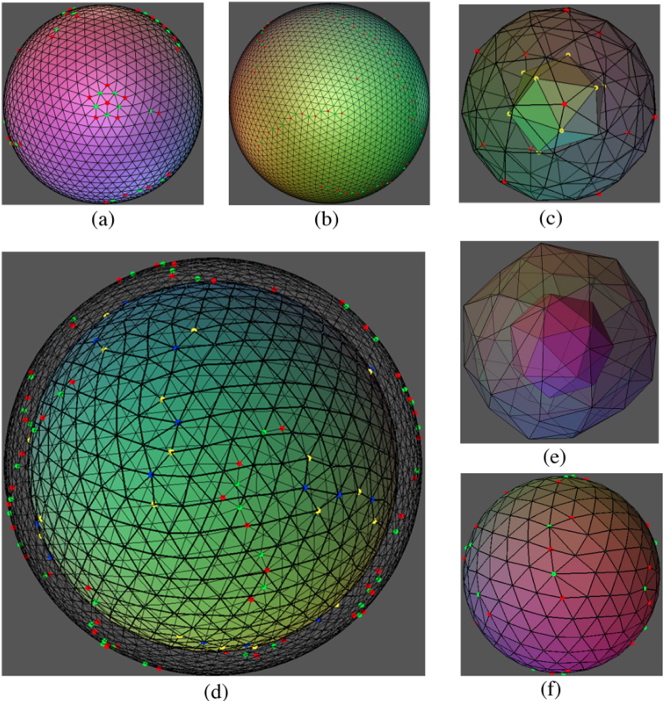}
\caption{Colour online. (a) Low energy metastable state of 2000 charges interacting via an inverse potential $\gamma=1$ (i.e. the Coulomb potential) showing an isolated rosette defect. (b) More typically however are grain boundary scars (of alternating positive and negative disclinations) as can be seen in this low energy metastable state of 5000 charges with $\gamma=1$. Positive and negative disclinations in (a) and (b) are coloured red and green respectively. (c ) A small cluster of 100 charges with $\gamma=2$; the particles are arranged into two distinct shells, an inner shell of 13 particles and an outer shell with 87 charges. (d) The two outermost shells for a system with 5000 charges and $\gamma=2$. Positive and negative disclinations on the inner shells in (c) and (d) are coloured yellow and blue, respectively.  (e) The two innermost shells in a cluster of 2000 charges with $\gamma=2$.  (f) Outer shell of a low energy state of 1000 charges interacting via an inverse potential with $\gamma=12$.}
\label{gallery}
\end{center}
\end{figure*}

Setting $\gamma=1$ in Eq. (\ref{eq:totalenergy}), we readily find a close correspondence between Eq. (\ref{eq:delta}) and our numerical results for N=1000, 2000 (fig.~\ref{gallery}a) and 5000 (fig.~\ref{gallery}b). These show that all the charges are indeed located in a single shell at the edge of the system. Using the Delaunay triangulation package Qhull we identify the number of nearest neighbours for each particle. Particles with five/seven nearest neighbours are coloured red/green, while particles with six neighbours are not highlighted. 

Euler's theorem stipulates that such spherical crystal cannot consist entirely of six coordinated particles but must also include a minimum of twelve pentagonal sites. Such points with an anomalous coordination are topological defects known as disclinations. Using Euler's theorem we can assign a topological charge to each disinclination, the sign and magnitude of the charge depends on how much the coordination number differs from 6. Thus, a pentagon has a topological charge +1 while a square has a charge of +2. Similarly a heptagon has a topological charge of -1 while an octagon has -2. Obviously a hexagon is topologically neutral. The total topological charge for any spherical cluster is conserved and must always be equal to +12 \cite{Bausch:2002}.

Such disclinations induce an enormous elastic strain in the lattice which can be reduced by arranging them symmetrically over the surface of the sphere. In addition the lattice may also include dislocations (tightly bound five-seven coordinated disclination pairs). Unlike disclinations, the number of dislocations is not fixed by topology and are only present if it is energetically favourable. 

Large clusters (those with more that 520 particles \cite{Wales:2009})  always contain dislocations. Typically, such dislocations condense around disclinations to form extended grain boundary ``scars'' of alternating positive-negative disclinations, as can be seen in the case for N=5000 (see fig.~\ref{gallery}b). The net topological charge of these scars is +1. More exotic disinclination structures, such as rosettes are also possible. An example is shown in the low energy state of a cluster of 2000 charges, see fig.~\ref{gallery}a. Such defects have a net topological charge of +1 and consist of a central positive disclination surrounded by five negative disclinations alternating with five positive disclinations. Rosette arrangements have been observed in both spherical \cite{Wales:2009} and flat crystals \cite{Radzvilavicius:2011}.

\section{ Second case: $\gamma=2$}

In the range $1<\gamma <3$ the charge density is expected to be intermediate between being entirely concentrated at the boundary and being uniform throughout the sphere. We concentrate on  $\gamma=2$ since in this case it is possible to derive the charge density in closed form.

\begin{figure}
\begin{center}
\includegraphics[width=1.0\columnwidth ]{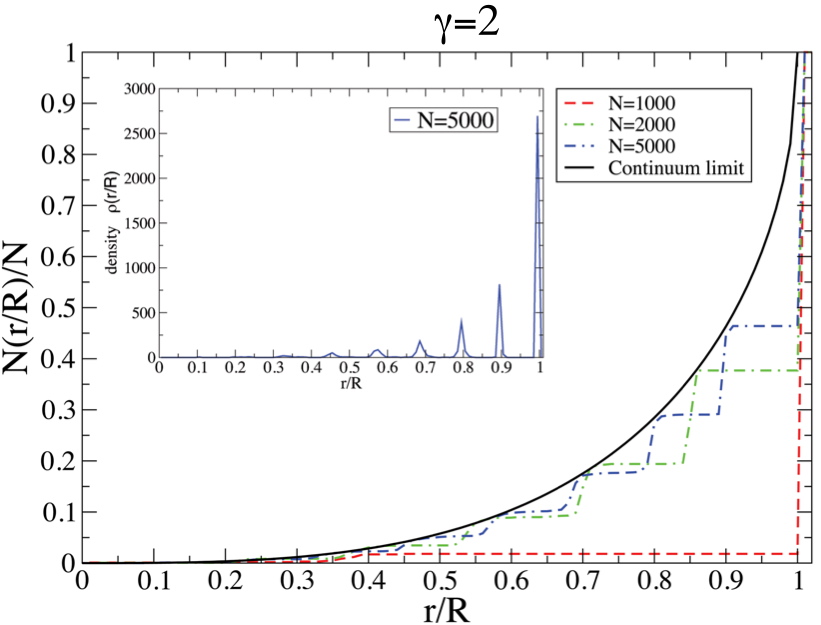}
\caption{Histograms of the fraction of the total charge of the system within a given radius r/R for $\gamma=2$. Results for clusters containing 1000, 2000, and 5000 charges are coloured red, green and blue respectively. The continuum limit result is given by the black dashed line. The inset shows the charge density for the system with 5000 particles.}
\label{cwrap}
\end{center}
\end{figure}

For $\gamma=2$ the energy given by Eq. (\ref{eq:totalenergy}) can be approximated by the integrals over the sphere $r\le R$,
\begin{equation}
E
=
\frac{1}{2}
\int\!\,d^3{r'}\int\!\,d^3{r}
\frac
{
\rho{({\bf r'})}\rho{({\bf r})}
}
{
|{\bf r}-{\bf r'}|^{2}
}
.
\label{eq:energy}
\end{equation}
The continuum approximation treats the density $\rho({\bf r})$ as a 
smooth function rather than the sum of delta functions,
\begin{equation}
\rho({\bf r})= \sum_{i=1}^{N}\delta({\bf r}-{\bf r}_i),
\label{eq:rhodef}
\end{equation}
where ${\bf r}_i$ is the position of the $i^{th}$ charge.
One then minimizes the energy of the cluster, with respect to the smooth function 
$\rho{({\bf r})}$, subject to the constraint that the number of particles,
\begin{equation}
N=\int\!\,d^3{r}\rho{({\bf r})},
\label{eq:Nparticles}
\end{equation}
is constant. Introducing the Lagrange multiplier $\mu$ the
constrained equation is, 
\begin{equation}
E
=
\int
\rho{({\bf r'})}
\left[
\frac{1}{2}
\int\!\,d^3{r}
\frac
{\rho{({\bf r})}}
{|{\bf r}-{\bf r'}|^{2}}
-
\mu
\right]
d^3{r'}
.
\end{equation}
A variation in the energy is given by,
\begin{equation}
\delta E = E{[\rho{({\bf r})} +\delta\rho{({\bf r})}]}-E{[\rho{({\bf r})}]},
\label{eq:Evariation}
\end{equation}
where $\delta\rho{({\bf r})}$ represents a small change in the charge
density. Keeping only terms up to first order,
Eq. ($\!\!$~\ref{eq:Evariation}) gives,
\begin{equation}
\delta E
=
\int\!\,
\delta\rho{({\bf r'})}
\left[
\int\!\,d^3{r}
\frac
{\rho{({\bf r})}}
{|{\bf r}-{\bf r'}|^{2}}
- 
\mu
\right]
d^3{r'}
,
\end{equation}
where to make the functional derivative stationary we require that,
\begin{equation}
\mu
=
\int\!\,d{^3}r'\frac{\rho{(r')}}{|{\bf r} - {\bf r'}|^{2}}.
\label{eq:n=1problem}
\end{equation}
Assuming spherical symmetry and expanding the denominator in the above equation as a power series, in terms of Legendre polynomials, yields, 
\begin{eqnarray}
\mu
&=&
2\pi
\int_{0}^{R}\!\,
\rho (r')
{r'}^2 dr'
\int_{0}^{\pi}\!\,d\theta
\sin \theta
\label{eq:Lseries}
\\
&&
\left[
\frac{1}{r_{>}^2}\sum_{n=0}^{\infty}\sum_{m=0}^{\infty}
\left(
\frac{r_{<}}{r_{>}}
\right)^{n+m} 
P_{n}(\cos\theta)P_{m}(\cos\theta)
\right] 
\nonumber
\end{eqnarray}
where, 
\begin{equation}
\cos \theta=\frac{ {\bf r}\cdot{\bf r'}}{|{\bf r}||{\bf r'}|},
\end{equation}
and to ensure convergence $r_{>}$ is the greater of $r$ and $r'$. Upon making the substitution $x=\cos\theta$ we can write Eq. ($\!\!$~\ref{eq:Lseries}) as,
\begin{eqnarray}
\frac{\mu}{2\pi}
&=&
\sum_{n=0}^{\infty}
\sum_{m=0}^{\infty}
\int_{0}^{R}\!\,
dr'
\rho(r')
\frac{{r'}^2}{r_{>}^2}
\left(
\frac{r_{<}}{r_{>}}
\right)^{m+n}
\\
\nonumber
&&
\int_{-1}^{1}\!\,dx
P_{n}(x) P_{m}(x),
\end{eqnarray}
and upon using the orthogonality conditions of Legendre polynomials we have,
\begin{equation}
\frac{\mu}{2\pi}
=
\sum_{n=0}^{\infty}
\frac{4\pi}{2n+1}
\int_{0}^{R}\!\,
dr'
\rho(r')
\frac{{r'}^2}{r_{>}^2}
\left(
\frac{r_{<}}{r_{>}}
\right)^{2n}.
\label{eq:pseries_form}
\end{equation}
Eq. ($\!\!$~\ref{eq:pseries_form}) can be split into two parts $r > r'$ and $r < r'$ giving,
\begin{eqnarray}
\frac{\mu}{2\pi}
&=&
\sum_{n=0}^{\infty}
\frac{4\pi}{2n+1}
\left[
\int_{0}^{r}\!\, dr' 
\rho (r')
\left( \frac{r'}{r}\right)^{2n+2}
\right.
\nonumber
\\
&&
\left.
+
\int_{r}^{R}\!\, dr' 
\rho (r')
\left( \frac{r}{r'}\right)^{2n}
\right],
\label{eq:another_pseries_form}
\end{eqnarray}
and writing the power series in Eq. ($\!\!$~\ref{eq:another_pseries_form})  in closed form we have,
\begin{eqnarray}
\frac{\mu r}{2\pi}
&=&
\int_{0}^{r}\!\, dr' g(r')r' \ln\left(\frac{r+r'}{r-r'} \right)
\nonumber 
\\
&&
-
\int_{r}^{R}\!\, dr' g(r')r' \ln\left( \frac{r+r'}{r-r'} \right ), 
\label{eq:log_closed_form}
\end{eqnarray}
where $g(r') = r'\rho(r')$. Differentiating both sides of Eq. ($\!\!$~\ref{eq:log_closed_form}) with respect to r yields,
\begin{eqnarray}
\frac{\mu }{2\pi}
&=&
\int_{0}^{R}\!\, dr' g(r')\frac{1}{r+r'} - \int_{0}^{r}\!\, dr' g(r')\frac{1}{r-r'}
\nonumber
\\
&&
\int_{r}^{R}\!\, dr' g(r')\frac{1}{r'-r}.
\label{eq:final_closed_form}
\end{eqnarray}
Treating the last two integrals in Eq. ($\!\!$~\ref{eq:final_closed_form}) as a Cauchy principle value integral we finally have,
\begin{equation}
\frac{\mu}{2\pi}
=
\int_{0}^{R}\!\,  dr' g(r') \frac{2r'}{r'^2 - r^2}.
\label{eq:final_closed_form_2}
\end{equation}
Eq. ($\!\!$~\ref{eq:final_closed_form_2}) is a singular integral equation of the first kind which can be solved to give \cite{Polyanin:2008},
\begin{equation}
\rho (r)
=
\frac{\mu}{2\pi^2}
\left(
\frac{1}{R^2-r^2}
\right)^{\frac{1}{2}}.
\label{eq:first_density}
\end{equation}
To find the Lagrange multiplier we substitute the Eq. ($\!\!$~\ref{eq:first_density}) into Eq. ($\!\!$~\ref{eq:Nparticles}) and solve for $\mu$, finally we have that the particle density is,
\begin{equation}
\rho (r)
=
\frac{N}{\pi^2}
\frac{1}{R^3(1-\frac{r}{R}^2)^{\frac{1}{2}}}
\label{eq:final_density}
\end{equation}
Thus in the continuum limit, unlike the case for $\gamma=1$, not all of the charge is found at the edge of the system. The fraction of charges within a fractional distance $r/R$ from the centre of the bounding sphere is given by integrating Eq. ($\!\!$~\ref{eq:final_density}), giving
\begin{equation}
\frac{N(\frac{r}{R})}{N}
=
\frac{2}{\pi}
\left(
\sin^{-1}\frac{r}{R}
-
\frac{r}{R}
\sqrt{
1+
\left(
\frac{r}{R}
\right)^2
}
\right)
\label{eq:enclosed_charges}
\end{equation}
Eq. ($\!\!$~\ref{eq:enclosed_charges}) can be compared with numerical results for low energy clusters with N=1000, 2000 and 5000 charges, see fig.~\ref{cwrap}. As expected, the agreement between numerical and analytical results improves with increasing N. 

In the case of such finite sized clusters the distribution shown in fig.~\ref{cwrap} displays a step like behaviour, particularly close to the edge of the system, which indicates that the charges form a series of concentric shells around the centre of the spherical box. This corresponds to a series of sharp well defined peaks of increasing density as shown in the inset in fig.~\ref{cwrap} for the N=5000 system. 

The morphology of small clusters $(N\lessapprox 500)$ in particular is dominated by the spherical hard wall boundary. In small systems all of the charges are arranged into well defined shells, where Euler's theorem holds individually for each shell. An example of a system with 100 charges is shown in fig.~\ref{gallery}c. 

For larger systems this is only true close to the edge of the system where the spherical boundary forces the charges to be concentrated into shells, corresponding to sharp peaks in density. Fig.~\ref{gallery}d shows the two outermost layers for N=5000. In both shells twelve grain boundary scars comprised of alternating positive and negative disclinations can be observed. However, towards the centre of the system the influence of the spherical boundary diminishes and the peaks in density become broader and less well defined. Consequentially, the shells in the inner region are irregularly shaped and it becomes difficult to uniquely identify distinct shells, see for example fig.~\ref{gallery}e which shows the two innermost shells for N=2000. This task is made more difficult by the presence of numerous isolated inter-shell charges in the interior region, i.e. charges which are found between adjacent shells and cannot be said to belong to either shell.

\section{ Third case: $\gamma=12$}

\begin{figure}
\begin{center}
\includegraphics[width=1.0\columnwidth ]{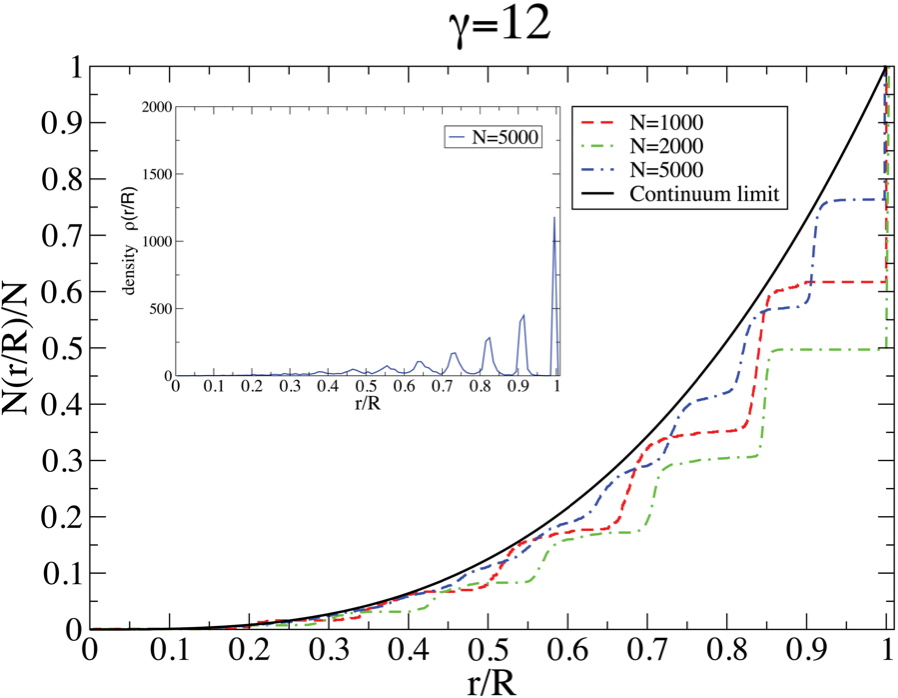}
\caption{Histograms of the fraction of the total charge of the system within a given radius r/R for $\gamma=12$. Results for clusters containing 1000, 2000, and 5000 charges are coloured red, green and blue respectively. The continuum limit result is given by the black dashed line. The inset shows the charge density for the system with 5000 particles.}
\label{cbbb}
\end{center}
\end{figure}

For $\gamma \geq 3$ the charge density in the continuum limit is expected to be uniform \cite{Hardin:2004}. We focus on the case of $\gamma=12$ which is the repulsive part of the familiar Lennard-Jones potential. Once again, we plot the fraction of charges $N(r/R)/N$, within a fractional radius $r/R$ and compare with the continuum limit result, see fig.~\ref{cbbb}. We find an improving agreement between analytical and numerical results. 

However, it is clear that in finite clusters there exists a systematic deviation from the expected uniform charge density, as seen by the fact that the peaks in density are much higher towards the edge of the system, see inset in fig.~\ref{cbbb} for N=5000. Such deviations are to expected (see \cite{Mughal:2008} for a comparative example) and a closer match to the charge distribution in finite clusters can be provided by higher order correction to the density that takes into account the shell like structure of the system close to the spherical boundary. It is possible that further annealing may yield a more uniform charge distribution but this has so far not proved to be the case, despite extensive numerical efforts. 

Again, near the spherical boundary the charges form a series of concentric spherical crystals, as an example the outermost layer in a system of 1000 charges is shown in fig.~\ref{gallery}f. However, towards the centre these shells are increasingly deformed, and it becomes difficult to uniquely identify separate shells (inter shell charges are also observed in the interior region). 

\section{Conclusions}

We studied a variant of the Thomson problem in which we seek the minimal energy arrangement of generalised charges (interacting via an inverse power law potential with exponent $\gamma$) in a spherical box. Increasing the value of $\gamma$ drives charges from the edge of the box into the interior. We find that in finite sized clusters the charges close to the edge of the spherical box are arranged into a series of concentric shells, within which the charges form well defined spherical crystals. However, towards the interior of the box the influence of the boundary is diminished and charges are arranged into less well defined configurations.

\acknowledgments
AM acknowledges useful discussions with Mike Moore.


\end{document}